\documentclass[floatfix,preprint, onecolumn, showpacs,nofootinbib]{revtex4}
\usepackage{natbib}
\usepackage{times}
\usepackage{amssymb,amsbsy,amsmath,amsfonts}
\usepackage{graphicx}

\usepackage{float}
\usepackage{morefloats}
\usepackage{rotating}
\usepackage{srcltx}
\usepackage{slashed}
\usepackage{subfigure}
\usepackage{color}

\begin{document}

\title{Chiral extrapolation and finite-volume dependence of the hyperon vector couplings}

\author{ Li-Sheng Geng,$^{1}$  Kai-Wen Li,$^{1}$ and J.~Martin Camalich,$^{3,4}$}
\affiliation{
$^1$School of Physics and Nuclear Energy Engineering and International Research Center for Particles and Nuclei in the Cosmos, Beihang University,  Beijing 100191,  China\\
$^3$Dept. Physics, University of California, San Diego, 9500 Gilman Drive, La Jolla, CA 92093-0319, USA\\
$^4$PRISMA Cluster of Excellence Institut f\"ur Kernphysik, Johannes Gutenberg-Universit\"at Mainz, 55128 Mainz, Germany}

\begin{abstract}
The hyperon vector form factors at zero momentum transfer, $f_1(0)$, play an important role in a precise determination
of the Cabibbo-Kobayashi-Maskawa matrix element $V_{us}$. Recent studies based on lattice chromodynamics (LQCD)  simulations and covariant baryon chiral perturbation theory yield contradicting results. In this work, we study chiral extrapolation of and finite-volume corrections to the latest $n_f=2+1$ LQCD simulations. Our results show that finite-volume corrections are relatively small and can be safely ignored at the present LQCD setup of $m_\pi L=4.6$ but chiral extrapolation needs to be performed more carefully. Nevertheless, the discrepancy remains and further studies are needed to fully understand it.
\end{abstract}


\maketitle

\section{Introduction}

The experimental determination of the Cabibbo-Kobayashi-Maskawa (CKM) matrix elements is of utmost importance for testing the flavor structure in the quark sector of the standard model~\cite{Antonelli:2009ws}. In particular, the elements of its first row provides a stringent test of the CKM unitarity~\cite{Cirigliano:2009wk}, namely $|V_{ud}|^2+|V_{us}|^2+|V_{ub}|^2=1$. With $|V_{ub}|$ in the ballpark of $10^{-3}$~\cite{Beringer:1900zz} and $|V_{ud}|=0.97425(22)$~\cite{Hardy:2008gy} precisely extracted from superallowed $0^+\rightarrow 0^+$ nuclear $\beta$ decays, the fulfillment of this constraint largely rests upon the determination of $|V_{us}|$.

The most precise value to date is provided by the analysis of (semi)leptonic kaon decays, which crucially depends on the accuracy at which
$f_K/f_\pi$ and $f_+(0)$  are known~\cite{Antonelli:2010yf}. The latest and remarkably precise lattice QCD (LQCD) computations of these quantities~\cite{Aoki:2013ldr,Bazavov:2012cd,Lubicz:2010bv} yield $|V_{us}|=0.2252(9)$. With this value, the first-row CKM unitarity turns out to be fulfilled at the permillage level~\cite{Beringer:1900zz,Aoki:2013ldr}.

Inclusive $\tau$ decays offer a completely independent extraction of this matrix element, yielding $|V_{us}|=0.2207(25)$~\cite{Gamiz:2004ar,Pich:2013lsa}, which is in slight tension with the kaon-decay determination and the CKM unitarity.

A third method to obtain $|V_{us}|$ is by studying semileptonic hyperon decays (for reviews see Refs.~\cite{Garcia:1985xz,Cabibbo:2003cu}). These are phenomenologically richer than their analogous kaon modes, in terms of multiple channels and polarization observables. However they are also considerably more complicated and up to six form factors can contribute per decay channel. At leading order in SU(3)-breaking, only two of these form factors evaluated at $q^2=0$ contribute, which are denoted as the vector and the axial hyperon couplings, $f_1(0)$ and $g_1(0)$. Furthermore, $f_1(0)$ is determined by conservation of the vector current up to $\mathcal{O}(m_s-m_{ud})^2$ corrections due to the Ademollo-Gatto theorem (AGt)~\cite{Ademollo:1964sr},\footnote{The $m_s$ and $m_{ud}$ generically denote the strange and the average $u$ and $d$ quark masses throughout this paper.} while the ratio $g_1(0)/f_1(0)$ can be obtained from an analysis of the angular dependence of the decay rates~\cite{Garcia:1985xz}. Reasoning along these lines and including only leading-order SU(3)-breaking corrections, Cabibbo and collaborators studied the hyperon semileptonic decay data and obtained $|V_{us}|=0.2250(27)$~\cite{Cabibbo:2003ea,Cabibbo:2003cu}, which is in perfect agreement with those determined from the kaon decays and the CKM unitarity.

However, this result does not include any estimate of the uncertainty produced by subleading SU(3)-breaking effects. In particular, it has been shown that an accurate knowledge of second-order breaking corrections to $f_1(0)$ is crucial to obtain a precise value of $|V_{us}|$~\cite{Mateu:2005wi}. Over the years various methods have been explored to calculate $f_1(0)$, including different quark models~\cite{Donoghue:1986th,Schlumpf:1994fb,Faessler:2008ix}, the MIT bag model~\cite{Donoghue:1981uk}, the large $N_c$ approach~\cite{FloresMendieta:1998ii,FloresMendieta:2004sk}, baryon chiral perturbation theory (BChPT)~\cite{Krause:1990xc,Anderson:1993as,Kaiser:2001yc,Villadoro:2006nj,Lacour:2007wm,Geng:2009ik} and quenched~\cite{Guadagnoli:2006gj,Lin:2008rb,Sasaki:2008ha,Gockeler:2011se} or $n_f=2+1$ LQCD~\cite{Sasaki:2012ne,Cooke:2012xv} simulations. As summarized in Refs.~\cite{Geng:2009ik,Sasaki:2012ne}, a puzzling outcome of these theoretical/numerical calculations is that the sign of the SU(3)-breaking corrections to $f_1(0)$ predicted in BChPT is  opposite to those found in most of the quark models and in LQCD.\footnote{For a quark model calculation predicting corrections to $f_1(0)$ of the same sign as BChPT see Ref.~\cite{Faessler:2008ix}. Fits to the semileptonic hyperon decay data using the Large $N_c$ parameterizations of the form factors~\cite{FloresMendieta:1998ii,FloresMendieta:2004sk} report the same sign too but the capacity of this approach to disentangle $V_{us}$ from the SU(3)-breaking corrections to $f_1(0)$ has been questioned~\cite{Mateu:2005wi}.}

The tension between the predictions of  BChPT and those of the LQCD simulations is particularly intriguing. On one hand, chiral perturbation theory is an effective field theory of QCD which relies on a perturbative expansion of its Green-functions about $p/\Lambda\sim$0, where $p$ is a small momenta or a light quark mass and $\Lambda\sim1$ GeV~\cite{ChiralIntro} (for a recent pedagogical review, see Ref.~\cite{Scherer:2012aa}). Its extension to the one-baryon sector is afflicted by the so-called power-counting-breaking (PCB) problem~\cite{Gasser:1987rb}. This can be solved by either implementing a non-relativistic expansion of the baryon fields, like in heavy-baryon ChPT~\cite{Jenkins:1990jv,Jenkins:1991es}, or keeping the theory relativistic while exploiting the fact that all PCBs are analytical and, therefore, they just affect the renormalization of low-energy constants (LECs) and not the physical results. Two renormalization prescriptions stand out among the manifestly covariant formalisms, the infrared (IR) ChPT~\cite{Becher:1999he} and the extended-on-mass-shell (EOMS) ChPT~\cite{Gegelia:1999gf,Fuchs:2003qc}. Although these approaches only differ in the organization of terms which are putatively of higher order, it has been shown in various phenomenological applications that EOMS ChPT tends to provide a faster convergence of the chiral series, especially in the three-flavor sector~\cite{Geng:2008mf,Geng:2009hh,MartinCamalich:2010fp} (see Ref.~\cite{Geng:2013xn} for a recent and comprehensive review).

On the other hand, LQCD simulations provide an \textit{ab initio} numerical solution of QCD from first principles in a finite hypercube, which can determine any nonperturbative matrix elements, such as $f_1(0)$, in a model-independent way. However, their very nature implies that simulations are performed at finite volume $T L^3$ with $T$ and $L$ the temporal and spatial extensions of the hypercube, and finite lattice spacing $a$. Furthermore, because of limitations in computing resources, most present LQCD simulations are performed at larger-than-physical light-quark masses (for a review see Ref.~\cite{Gattringer:2010zz}). Therefore, to obtain the physical quantities, extrapolations in terms of light-quark masses, often termed as chiral extrapolation, lattice volume and lattice spacing are necessary. In fact, a precise quark mass dependence and finite-volume effects are known to play an important role in many physical observables simulated on the lattice, such as baryon masses~\cite{Ali Khan:2003cu,Procura:2006bj,Young:2009zb,MartinCamalich:2010fp,Semke:2011ez,Geng:2011wq,Ren:2012aj,Alvarez-Ruso:2013fza}, magnetic moments and charge radii~\cite{Beane:2004tw,Tiburzi:2007ep,Jiang:2008ja,Wang:2008vb,Ledwig:2011cx,Greil:2011aa,Wang:2012hj,Hall:2013oga}, the nucleon axial charge~\cite{Beane:2004rf,Procura:2006gq,Khan:2006de} and the electromagnetic/vector current of the nucleon~\cite{Tiburzi:2007ep,Jiang:2008ja,Ledwig:2011mn,Greil:2011aa}. In particular, finite-volume corrections (FVCs) are believed to be responsible for the discrepancy between the LQCD simulated $g_1$ and its experimental counterpart~\cite{Lin:2012ev}.

Given the fact that a precise $f_1(0)$ is of ultimate importance to the extraction of $V_{us}$ from hyperon decays, we study in this work the chiral extrapolation of and finite-volume corrections to the hyperon vector couplings in BChPT. In particular, we will investigate whether these effects can explain the discrepancies between the BChPT and LQCD predictions. This article is organized as follows. In Sec. 2, we recall the computation of $f_1(0)$ up to $\mathcal{O}(p^4)$ in the continuum, the implication of the AGt and its caveat. We then explain how light-quark mass dependence of $f_1(0)$ is determined and present for the first time the formalism to calculate finite-volume corrections. In Sec. 3, we formulate ChPT in finite volume and calculate numerically the FVCs to the LQCD simulations of Ref.~\cite{Sasaki:2012ne}. A short summary is given in Sec. 4.

\section{The hyperon vector coupling in BChPT}\label{sec:BChPT}

The baryon vector form factors as probed by the charged $\Delta$S=1 weak current
$V^\mu=V_{us}\bar{u}\gamma^\mu s$
are defined by
\begin{equation}\label{eq:f1}
\langle B_2\vert V^\mu\vert B_1\rangle =V_{us}\bar{u}(p')\left[\gamma^\mu f_1(q^2)+\frac{i \sigma^{\mu\nu}q_\nu}{M_{1}+M_2}f_2(q^2)+\frac{q^\mu }{M_{1}+M_2}f_3(q^2)\right]u(p),
\end{equation}
where $q=p_2-p_1$. The properties of the three form factors, $f_1$, $f_2$, and $f_3$, can be found in Ref.~\cite{Cabibbo:2003cu}. The chiral corrections to the hyperon vector coupling, $f_1(0)$, can be parameterized order-by-order as,
\begin{eqnarray}
&&f_1(0)=g_V(1+\delta f_1),\nonumber\\
&&\delta f_1=\delta^{(2)}+\delta^{(3)}+\cdots,
\end{eqnarray}
where, in consistency with previous calculations~\cite{Villadoro:2006nj,Lacour:2007wm,Geng:2009ik}, we have denoted the $\mathcal{O}(p^3)$ and $\mathcal{O}(p^4)$ chiral corrections by $\delta^{(2)}$ and $\delta^{(3)}$, respectively. The vector couplings are fixed in the SU(3)-symmetric limit by $g_V=-\sqrt{3/2}$, $-\sqrt{1/2}$, $-1$, $\sqrt{3/2}$, $\sqrt{1/2}$, and $1$ for $\Lambda\rightarrow p$, $\Sigma^0\rightarrow p$, $\Sigma^-\rightarrow n$, $\Xi^-\rightarrow \Lambda$, $\Xi^-\rightarrow\Sigma^0$, and $\Xi^0\rightarrow\Sigma^+$.  In the isospin-symmetric limit, only four channels provide independent information, which are $\Lambda N$, $\Sigma N$, $\Xi \Lambda$, and $\Xi\Sigma$.

The chiral expansion of $f_1(0)$ has some features worth reminding here. The first one is an important caveat concerning the AGt in the context of spontaneous chiral symmetry breaking~\cite{Langacker:1973nf}. It is well known that the leading chiral loop corrections to the hyperon (and kaon) vector couplings scale as $\sim(m_s-m_{ud})^2/(m_s+m_{ud})$~\cite{Gasser:1984ux,Anderson:1993as}, which literally fulfills the suppression predicted by the AGt as long as $(m_s-m_{ud})\ll(m_s+m_{ud})$. However, in the physical world $m_{ud}\ll m_s$ and the chiral loops with virtual octet baryons are expected to scale as $\delta^{(2)}\sim\mathcal{O}(m_s)$ and $\delta^{(3)}\sim\mathcal{O}(m_s^{3/2})$~\cite{Anderson:1993as}. The contributions of virtual decuplets to $f_1(0)$ are more complicated due to the inclusion of the octet-decuplet mass splitting $\Delta$, which is a small parameter in the approach that does not vanish in the chiral or the SU(3)-symmetric limit. The chiral loops with decuplet baryons also fulfill the AGt theorem explicitly~\cite{Villadoro:2006nj}, but their actual behavior at $m_{ud}\ll m_s$ turns out to be $\mathcal{O}(\Delta\times m_s^{1/2})$ modulo a nonanalytical function of $m_s$ and $\Delta$.

The first analytical corrections to $f_1(0)$ start at $\mathcal{O}(p^5)$ in the chiral expansion and they would scale as $\mathcal{O}(m_s^2)$. An important consequence of this is that there are no unknown LECs contributing to the chiral expansion of $\delta f_1(0)$ until this order~\cite{Krause:1990xc,Anderson:1993as}. Thus, up to $\mathcal{O}(p^4)$, BChPT is completely predictive in the determination of the SU(3)-breaking corrections to $f_1(0)$. There are no PCB terms and a study in the original covariant formulation of BChPT happens to be equivalent to the EOMS one~\cite{Geng:2009ik}. In the following we summarize this calculation in the covariant formalism including the decuplet baryons, placing especial emphasis on the quark mass dependence of the results.

\begin{figure}[t]
\center
\subfigure[]{
\includegraphics[scale=0.8,angle=270]{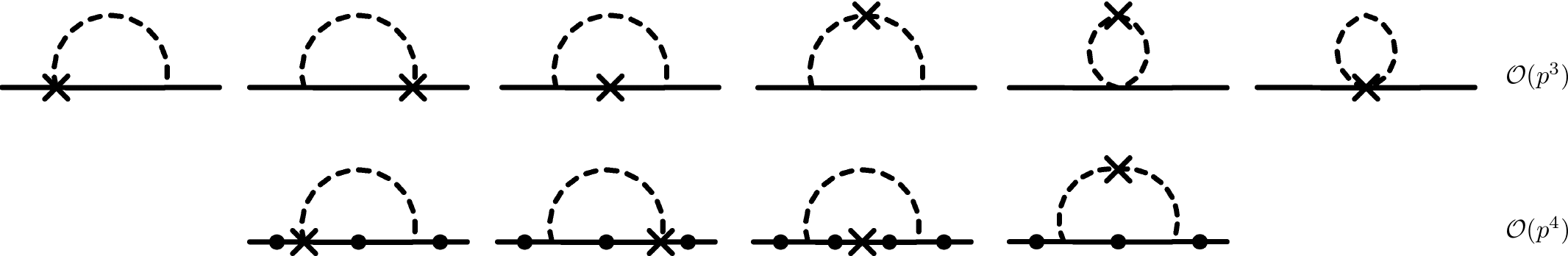}
}\\
\subfigure[]{
\includegraphics[scale=0.8,angle=270]{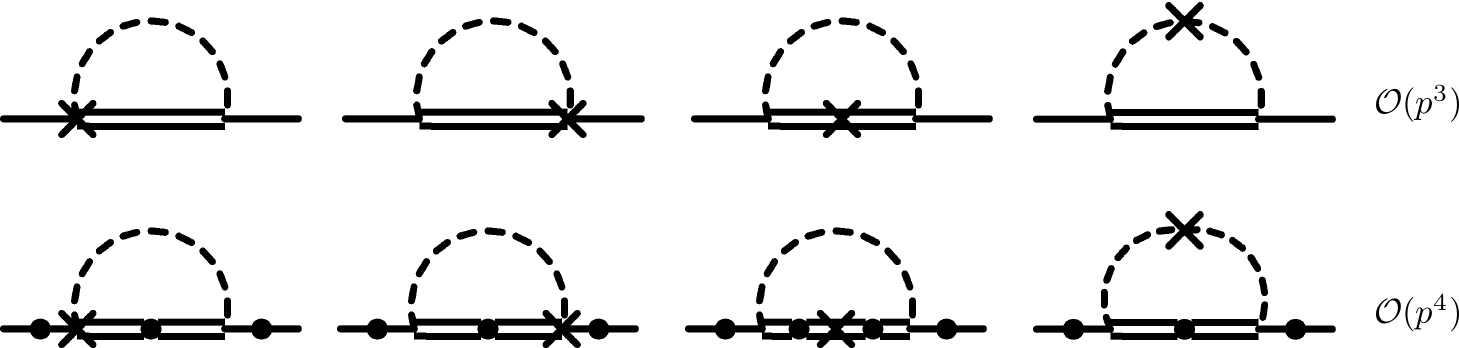}
}

\caption{Virtual octet (a) and decuplet contributions (b) to $f_1(0)$ up to $\mathcal{O}(p^4)$.  The solid lines correspond to octet baryons, double lines to decuplet baryons, and dashed lines indicate mesons; crosses indicate the coupling of the external current; black dots denote mass splitting
insertions. We have not shown explicitly those diagrams corresponding to wave function renormalization, which have been taken into account in the calculation. }\label{fig:diagrams}
\end{figure}

\subsection{Quark mass dependence of $f_1(0)$}

In Fig.~\ref{fig:diagrams} we show all the relevant Feynman diagrams that contribute to the chiral expansion up to $\mathcal{O}(p^4)$ and with the explicit inclusion of decuplet baryons. Note that wave-function renormalization must also be included. As discussed above, no unknown LECs contribute to the calculation up to this order and the BChPT prediction only depends on the values of the meson semileptonic decay constant $F_0$, the baryon axial couplings $D$, $F$ and $\mathcal{C}$, and the pseudoscalar meson and baryon masses (details can be found in Ref.~\cite{Geng:2009ik}). Up to $\mathcal{O}(p^4)$ it remains a good approximation to treat $F_0$ and the axial couplings as quark-mass independent parameters that we fix at their SU(3)-averaged physical values: $F_0=1.17\,F_\pi$, $D=0.80$, $F=0.46$~\cite{Cabibbo:2003cu} and $\mathcal{C}=0.85(5)$~\cite{Alarcon:2012nr}. The latter value is obtained using an average of the different hadronic decuplet decays while in our previous analysis~\cite{Geng:2009ik} we used  $\mathcal{C}=1.0$ which is obtained only from the $\Delta\rightarrow\pi N$ decay rate. The pion and kaon masses ultimately provide the source of SU(3)-breaking in the theory and they are adjusted to their physical values or to the ones obtained at the unphysical quark masses employed in the LQCD simulations. We obtain the $\eta$ mass using the Gell-Mann-Okubo mass formula, $m_\eta^2=(4m_K^2-m_\pi^2)/3$, which also holds in the $\eta$ contributions to $f_1(0)$ up to $\mathcal{O}(p^4)$.

In the calculation of $\delta^{(2)}$ one can work with the physical averages of the baryon-octet and -decuplet masses, $M_{B0}=1.151$ GeV and $M_{D0}=1.382$ GeV. The only corrections at $\mathcal{O}(p^4)$ actually stem from the baryon mass splittings entering the loop diagrams, contributions which are denoted by the dots in the diagrams of Fig.~\ref{fig:diagrams}~\cite{Villadoro:2006nj,Geng:2009ik}. Although these insertions could be performed at the perturbative level, in the present calculation they are implemented to all orders by including the SU(3)-symmetry broken masses in the propagators and in the on-shell conditions $p_1^2=M_1^2$ and $p_2^2=M_2^2$~\cite{Geng:2009ik}.

Up to $\mathcal{O}(p^4)$ in $f_1(0)$, or $\delta^{(3)}$, it suffices to work with the $\mathcal{O}(p^2)$ chiral formulas of the baryon masses, which depend on the four LECs $M_{B0}$, $b_0$, $b_D$ and $b_F$ for the octet baryons and the three LECs $M_{D0}$, $g_0$, $g_D$ for the decuplet baryons (we follow the notation and conventions of Ref.~\cite{MartinCamalich:2010fp}). These formulas reproduce accurately the experimental data and describe quite well the quark mass dependence of the baryon masses. As we discuss in the next section, we will make use of them to complete information on the baryon masses that is not provided by the LQCD analyses.

Finally, the results at $\mathcal{O}(p^4)$ contain higher-order divergences that are renormalized in the $\overline{MS}$ scheme. By setting the corresponding LECs to zero but studying the residual renormalization-scale dependence, $0.7$ GeV$\leq\mu\leq1.3$, we obtain an estimate of the systematic uncertainty of $\sim\mathcal{O}(m_s^2)$ due to the truncation of the chiral series at $\mathcal{O}(p^4)$~\cite{Geng:2009ik}.

\subsection{Finite-volume correction to $f_1(0)$}

In this section, we present the calculation of the FVCs to $f_1(0)$. The temporal extension in LQCD simulations is much larger than the spatial ones and we treat it as a continuous and infinite variable. The spatial components are contained within a three dimensional box with periodic boundary conditions and we assume that we work in the $p$-regime, $L\cdot m_\pi\gg1$~\cite{Gasser:1987zq}. These corrections can be calculated in BChPT using the same set of diagrams as in the continuum theory, Fig.~\ref{fig:diagrams}, and discretizing the spatial components of the 4-momentum loop integrals, $\int\limits_{-\infty}^{+\infty} d k^0 \sum\limits_{\vec{n}=-\infty}^\infty \left(\frac{2\pi}{L}\right)^3 \vec{n}$. Since one is now treating differently the temporal and spatial components of a loop diagram one breaks Lorentz invariance and the decomposition of the matrix element in Eq.~(\ref{eq:f1}) does not hold in this case.

A way to circumvent this problem in the determination of $f_1(0)$ is to calculate a scalar quantity that results from taking the divergence upon this matrix element~\cite{Guadagnoli:2006gj,Sasaki:2008ha}:
\begin{eqnarray}
q_\mu \langle B'\vert V^\mu\vert B\rangle =V_{us}(M_{2}-M_1)\bar{u}(p')\, f_S(q^2)  u(p),
\end{eqnarray}
where we have introduced the so-called scalar form factor,
\begin{equation}
f_S(q^2)=f_1(q^2)+\frac{q^2}{M_{2}^2-M_1^2} f_3(q^2).
\end{equation}
Therefore, the calculation of the hyperon vector coupling in the finite volume can be simplified by computing $f_S(q^2)$ and setting $q^2=0$. Furthermore, the calculation of the scalar form factor at (euclidean) maximum recoil $q^2_{E, max}=-(M_1-M_2)^2$ presents many numerical advantages in a LQCD computation and it can be obtained with high precision in the simulations~\cite{Guadagnoli:2006gj}. On the other hand, doing so requires an additional interpolation to $q^2=0$ helped by less precise results at (euclidean) $q_E^2>$0, which are obtained calculating the customary 3-point functions.

In principle, one would like to investigate FVCs to $f_S(q^2)$ at the (minkowskian) maximum recoil $q^2_{max}=(M_1-M_2)^2$ in BChPT and then use the resulting corrected results to interpolate to $q^2=0$. However this method has drawbacks since new terms beyond those shown in Fig.~\ref{fig:diagrams}, with unknown LECs, can contribute to $f_S(q^2)$ at $\mathcal{O}(p^4)$. Therefore, we choose to study the FVCs of the quantity $f_S(0)=f_1(0)$ by putting the initial baryon at rest, i.e., $p=(M_1,\vec{0})$. The condition of $q^2=0$ indicates the four-momentum of the final baryon to be $p'=(E_F,p_F,0,0)$~\footnote{Any other choice for the spatial three momenta will yield the same results because of the remaining cubic symmetry.} with $E_F=(M_{2}^2+M_1^2)/(2M_1)$ and $P_F=\sqrt{E_F^2-M_{2}^2}$. Following the procedures outlined in
Ref.~\cite{Geng:2011wq}, one can now easily calculate all the relevant loop diagrams in finite volume.

As recognized in Ref.~\cite{Geng:2009ik}, the $\mathcal{O}(p^4)$ results are rather lengthy and we refrain from writing them down explicitly. The $\mathcal{O}(p^3)$ results are quite compact and, for the sake of completeness, we present them in the Appendix.

\section{Results}

\subsection{Results at the physical point revisited}

\begin{table}[t]
\renewcommand{\arraystretch}{1.0}
\setlength{\tabcolsep}{0.2cm}
\caption{Results for the chiral corrections to $\delta f_1(0)$ (in percentile) up to $\mathcal{O}(p^4)$ in covariant BChPT and including the decuplet resonances as explicit degrees of freedom~\cite{Geng:2009ik}. We separate the results in $\mathcal{O}(p^3)$ and  $\mathcal{O}(p^4)$, $\delta^{(2)}$ and  $\delta^{(3)}$, respectively, and in the contributions given by virtual octets (O) or decuplets (D) in the loops.  \label{table:phys_results}}
\begin{tabular}{|c|cc|cc|c|}
\hline
&$\delta^{(2)}$ (O)&$\delta^{(2)}$ (D)& $\delta^{(3)}$ (O)&$\delta^{(3)}$ (D)&Total\\
\hline
$\Lambda\, N$&$-3.8$&+0.5&$+0.2^{+1.2}_{-0.9}$&$+2.2^{+0.1}_{-0.1}$&$-0.9^{+1.3}_{-1.0}$\\
$\Sigma\,N$&$-0.8$&$-1.0$&$+4.7^{+3.8}_{-2.8}$&$+4.5^{+0.3}_{-0.2}$&$+7.4^{+4.1}_{-3.0}$\\
$\Xi\Lambda$&$-2.9$&$-0.01$&$+1.7^{+2.4}_{-1.8}$&$+3.8^{+0.3}_{-0.2}$&$+2.6^{+2.7}_{-2.0}$\\
$\Xi\Sigma$&$-3.7$&+0.5&$-1.3^{+0.3}_{-0.2}$&$+4.3^{+1.4}_{-1.0}$&$-0.2^{+1.7}_{-1.2}$\\
\hline
\end{tabular}
\end{table}

In Table~\ref{table:phys_results} we list the results for the chiral corrections to $f_1(0)$ up to $\mathcal{O}(p^4)$ in covariant BChPT and including the decuplet resonances as explicit degrees of freedom. These values are an update with respect to those presented in Ref.~\cite{Geng:2009ik} and the differences originate from the slightly smaller $\mathcal{C}$ used in the current analysis. As it was already pointed out in Ref.~\cite{Geng:2009ik}, the corrections at $\mathcal{O}(p^4)$ are generally larger than those at $\mathcal{O}(p^3)$ and this seems to suggest that the chiral convergence for $f_1(0)$ is broken. Limiting ourselves to the octet contributions, a naive power-counting estimate of the potential size of these SU(3)-breaking corrections indicates that $\delta^{(2)}\sim \frac{m_K^2}{\Lambda^2}\sim20\%$ and $\delta^{(3)}\sim \frac{m_K\Delta_{12}}{\Lambda^2}\sim10\%$, with $\Delta_{12}=M_1-M_2$.~\footnote{Note that a similar argument can be made for the decuplet contributions by taking the limit $\Delta\rightarrow 0$.}
Therefore, it is difficult to judge the convergence of the chiral series of $f_1(0)$ by comparing the third and fourth orders in the expansion since, as shown in Table~\ref{table:phys_results}, the leading terms are suppressed by small coefficients and $\delta^{(2)}$ turns out to be roughly a factor ten smaller than the power-counting estimate ~\cite{Anderson:1993as}. A similar phenomenon is observed in the leading contributions to the kaon vector form factor~\cite{Gasser:1984ux}. Nonetheless, the BChPT results already contain an estimate of the higher-order uncertainty which comes from varying $\mathcal{O}(p^5)$ analytical pieces (renormalization scale dependence) and one sees that these can be sizable and as large as a few percent.

\subsection{Chiral extrapolation at $m_s\simeq m_{s,phys}$}

In the following, we study the light-quark mass dependence of $f_1(0)$  by analyzing the only $n_f=2+1$ LQCD results reported at the moment for the hyperon charges and in the channels $\Xi\Sigma$ and $\Sigma N$~\cite{Sasaki:2012ne}. These simulations are performed using RBC and UKQCD ensembles generated in a $24^3\times64$ grid with periodic boundary conditions in the spatial dimensions~\cite{Allton:2008pn}. The quarks are described by a domain wall fermion action (known to have improved chiral symmetry properties) and with the strange quark mass tuned to be approximately equal to the physical one. The lattice spacing is determined using the $\Omega^-$ mass, $a=0.114(2)$ fm, making the full length of the spatial extensions $L\sim2.736$ fm.

In Table~\ref{table:lattice_masses} we show the values of the meson and baryon masses for the different quark masses reported in Ref.~\cite{Sasaki:2012ne}. Errors are omitted because they have a negligible impact on the $f_1(0)$ results. The $\pi$, $K$, $N$, $\Sigma$ and $\Xi$ masses are determined and given in Ref.~\cite{Sasaki:2012ne}. For the $\Lambda$ baryon mass the $\mathcal{O}(p^2)$ formulas for the baryon masses are equivalent to the Gell-Mann-Okubo formula, $M_\Lambda=(2M_N+2M_\Xi-M_\Sigma)/3$, which is fulfilled experimentally very accurately and it seems to hold also for unphysical quark masses as those discussed here~\cite{Beane:2006pt}. For the quark mass dependence of the decuplet-baryon masses we have little information from the actual ensembles used in Refs.~\cite{Sasaki:2012ne,Allton:2008pn} and in this case we use the $\mathcal{O}(p^2)$ mass formulas with the LECs determined from  LQCD data~\cite{Ren:2013oaa}, $M_{D0}=1.135$ GeV, $\gamma_M=0.167$ GeV$^{-1}$ and $\gamma_M=0.322$ GeV$^{-1}$.

\begin{table}[t]
\renewcommand{\arraystretch}{1.0}
\setlength{\tabcolsep}{0.2cm}
\caption{Masses of the pseudoscalar mesons and the octet and decuplet baryons in units of GeV, determined as explained in the main text for the different ensembles employed in the $n_f=2+1$ LQCD simulations of Ref.~\cite{Sasaki:2012ne}. \label{table:lattice_masses}}
\begin{tabular}{|cc|cccc|cccc|}
\hline
$m_\pi$  &$m_K$  & $M_N$ & $M_\Lambda$ & $M_\Sigma$ & $M_\Xi$ & $M_\Delta$ & $M_{\Sigma^*}$ & $M_{\Xi^*}$ & $M_{\Omega^-}$\\
\hline
0.330&0.576&1.140 &1.271& 1.330&1.431&1.369&1.513&1.656&1.800\\
0.420&0.606&1.237&1.339&1.386&1.465&1.458&1.580&1.703&1.826\\
0.558&0.665&1.412&1.470&1.501&1.544&1.635&1.720&1.804&1.888\\
\hline
\end{tabular}
\end{table}

\begin{figure}[t]
\centerline{\includegraphics[scale=0.45]{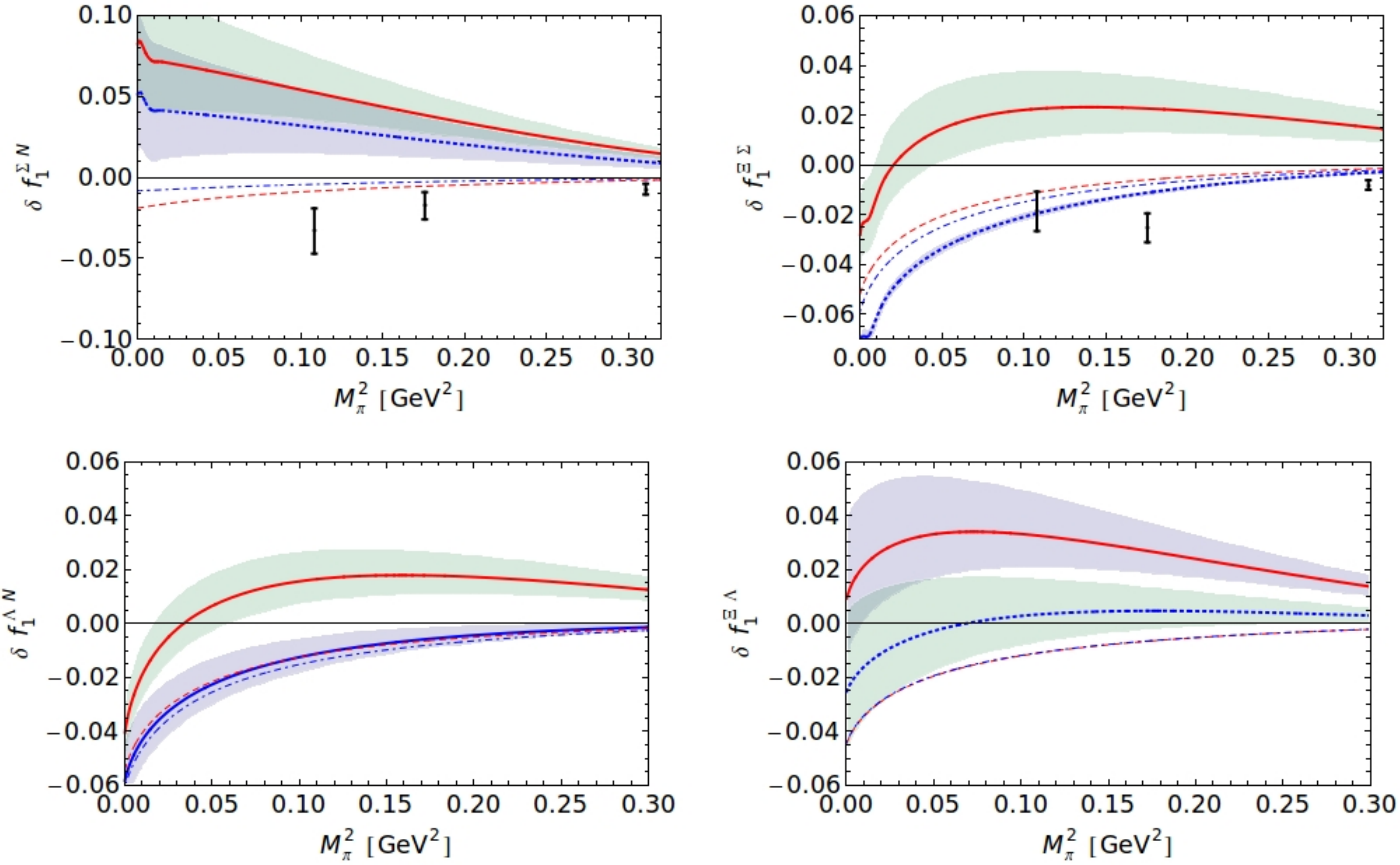}}
\caption{SU(3) breaking corrections to $f_1(0)$ in infinite volume as functions of the pion mass, $m_\pi$, in the different approaches of covariant BChPT discussed in the text. The (red) solid lines are the $\mathcal{O}(p^4)$ results including octet and decuplet contributions, the (blue) dotted lines are $\mathcal{O}(p^4)$ results including only the octet corrections, the (red) dashed lines are $\mathcal{O}(p^3)$ results including octet and decuplet contributions and the (blue) dashed-dotted lines are $\mathcal{O}(p^3)$ results including only the octet corrections.\label{fig:plotsasaki1}}
\end{figure}

In Tables~\ref{table:octetfull} and~\ref{table:octetdecupletfull} we tabulate the SU(3)-breaking corrections to $f_1(0)$ predicted by covariant BChPT at the simulated light-quark masses without and with decuplet degrees of freedom, respectively. In each of these two tables we include the $\mathcal{O}(p^3)$ and $\mathcal{O}(p^4)$ results, whereas the respective FVCs are given in the parentheses. In the last column we show the results extracted from the simulations~\cite{Sasaki:2012ne}. In Fig.~\ref{fig:plotsasaki1} we show the pion mass dependence of our results in the four channels and for the different cases compared against the LQCD points.\footnote{In order to describe the quark mass dependence of $f_1(0)$ in the plots,  we use phenomenological interpolators to reproduce accurately the pion-mass dependence of the kaon and baryon octet masses obtained from the LQCD configurations. In case of $M_\Lambda$ we always use the GMO relation, whereas for the decuplet we use the LO BChPT formulas.} It is important to note that the chiral corrections diminish as we approach the SU(3)-symmetric point at larger pion masses. However, the actual values at the two heavier masses should be interpreted with caution since these points are at the border or beyond the range of applicability of BChPT.

\begin{table}[t]
 \begin{center}
 \end{center}
      \renewcommand{\arraystretch}{1.0}
     \setlength{\tabcolsep}{0.2cm}
     \caption{Virtual octet contributions to the SU(3)-breaking corrections to $f_1(0)$ in covariant BChPT (in percentage). The uncertainties are
     obtained by varying $\mu$ from 0.7 to 1.3 GeV. Finite-volume correction are given in the parentheses\label{table:octetfull}}
\begin{tabular}{|c|ccc|c|}
\hline
                          &$\delta^{(2)}$&  $\delta^{(3)}$ & $\delta^{(2)}+\delta^{(3)}$& LQCD~\cite{Sasaki:2012ne}\\
			\hline
 $\Sigma N$  &  $-0.57(-0.11)$   &  $3.7_{-1.7}^{+2.3}(-0.08)$       & $3.1_{-1.7}^{+2.3}(-0.19)$       & $-3.44\pm1.4$ \\
                        &  $-0.37(-0.03)$   &  $2.5_{-1.0}^{+1.4}(-0.03)$       & $2.1_{-1.0}^{+1.4}(-0.06)$       & $-1.84\pm0.84$ \\
                        &  $-0.14(0.00)$    &  $1.0_{-0.3}^{+0.6} (0.00)$       & $0.8_{-0.3}^{+0.6}(0.00)$       & $-0.81\pm0.33$ \\
\hline
 $\Xi\Sigma$ &  $-1.58(0.09)$    &  $-0.5_{-0.1}^{+0.1} (0.00)$   & $-2.1_{-0.2}^{+0.1}(0.09)$       & $-1.92\pm0.79$  \\
                        &  $-0.85(0.03)$    &  $-0.3_{-0.1}^{+0.1} (0.00)$   & $-1.2_{-0.1}^{+0.1}(0.03)$       & $-2.58\pm0.58$\\
                        &  $-0.25(0.00)$    &  $-0.0_{-0.1}^{+0.1} (0.00)$   & $-0.3_{-0.1}^{+0.1}(0.00)$       & $-0.86\pm0.19$  \\
\hline
\end{tabular}
\end{table}

\begin{table}[t]
\begin{center}
 \end{center}
      \renewcommand{\arraystretch}{1.0}
     \setlength{\tabcolsep}{0.2cm}
     \caption{Virtual octet plus virtual decuplet contributions to the SU(3)-breaking corrections to $f_1(0)$ in covariant BChPT (in percentage). The uncertainties are
     obtained by varying $\mu$ from 0.7 to 1.3 GeV. Finite-volume correction are given in the parentheses. \label{table:octetdecupletfull}}
\begin{tabular}{|c|ccc|c|}
\hline
                         &$\delta^{(2)}$&  $\delta^{(3)}$ & $\delta^{(2)}+\delta^{(3)}$& LQCD~\cite{Sasaki:2012ne} \\
			\hline
 $\Sigma N$   & $-1.14(-0.07)$   &  $7.0_{-1.7}^{+2.3}(-0.09)$   & $5.9_{-1.7}^{+2.3}(-0.16)$      & $-3.44\pm1.4$  \\
                         & $-0.70(-0.02)$   &  $4.7_{-1.0}^{+1.4} (-0.03)$  & $4.0_{-1.0}^{+1.4}(-0.05)$     & $-1.84\pm0.84$ \\
                         & $-0.25(0.00)$    &  $1.8_{-0.2}^{+0.6} (-0.01)$  & $1.5_{-0.2}^{+0.6} (-0.01)$    & $-0.81\pm0.33$ \\
\hline
 $\Xi\Sigma$  & $-1.26 (0.06)$   &   $3.8_{-1.4}^{+1.8}(0.00)$  & $2.5_{-1.4}^{+1.8}  (0.06)$   & $-1.92\pm0.79$  \\
                         & $-0.66 (0.02)$   &   $2.9_{-0.9}^{+1.2}(0.00)$  & $2.2_{-0.9}^{+1.2}  (0.02)$    & $-2.58\pm0.58$  \\
                         & $-0.19 (0.00)$   &   $1.5_{-0.5}^{+0.6}(0.00)$  & $1.3_{-0.5}^{+0.6}  (0.00)$    & $-0.86\pm0.19$  \\
 \hline
\end{tabular}
\end{table}

The first thing worth noticing is that the BChPT results for the quark mass dependence of $f_1(0)$ depend very much on the order of the calculation or on the inclusion of the decuplet degrees of freedom. For instance, in the $\Sigma N$ channel our results at $\mathcal{O}(p^3)$ remain negative and small, even after accounting for the virtual decuplet contributions. The size predicted is smaller than that obtained in LQCD at this order. However, the corrections at $\mathcal{O}(p^4)$ are large and positive, making the net effect up to this order of about $\sim+5\%$, in stark contrast with LQCD, whose result is sizable but negative. For the $\Xi\Sigma$ channel the agreement with LQCD is better and it even improves at $\mathcal{O}(p^4)$ if the decuplet contributions at this order are not included. However, their inclusion pushes the total contribution to be positive also in this channel. One would hope that finite-volume corrections would account for the differences between BChPT and LQCD, but this is not the case. As it is shown by the values in parenthesis in  Tables~\ref{table:octetfull} and~\ref{table:octetdecupletfull}, these are very small and negligible at these quark masses simulated in the $(2.736)^3$ fm$^3$ lattices. Overall, an agreement between BChPT up to $\mathcal{O}(p^4)$ and the LQCD results for the $\Sigma N$ and $\Xi \Sigma$ channels~\cite{Sasaki:2012ne} is not apparent.

\begin{table}[t]
\begin{center}
 \end{center}
      \renewcommand{\arraystretch}{1.0}
     \setlength{\tabcolsep}{0.2cm}
     \caption{Results on $f_1(0)$ (in percentage) at the physical point using fits to LQCD points with covariant BChPT  up to $\mathcal{O}(p^4)$ plus an analytical piece of $\mathcal{O}(p^5)$. The first error is statistic and the second theoretical stemming from unknown $\mathcal{O}(p^5)$ pieces. We compare these with the results obtained using AGt inspired fits done in Ref.~\cite{Sasaki:2012ne}. \label{table:pheno_fit}}
\begin{tabular}{|c|cc|c|}
\hline
&$c_{12}$&Ch. 1-loop+LEC[$\mathcal{O}(p^5)$] &AGt~\cite{Sasaki:2012ne}\\
\hline
$\Sigma N$&$-1.40(12)$&$-0.6(0.7)(3.5)$&$-2.66(63)$ \\
$\Xi\Sigma$&$-1.16(8)$&$-6.6(0.4)(1.4)$&$-2.63(39)$  \\
 \hline
\end{tabular}
\end{table}

As explained above, one cannot deduce the breakdown of the chiral expansion from the comparison between $\delta^{(2)}$ and $\delta^{(3)}$, although enforcing an agreement between the  BChPT results and the current LQCD results would require large $\mathcal{O}(p^5)$ contributions. In order to quantify this, we add an analytical piece of $\mathcal{O}(p^5)$ to the chiral-loops,
\begin{equation}
 \delta^{(4)}=c_{12}(m_K^2-m_\pi^2)^2, \label{eq:AGtf1}
\end{equation}
and fit the constant to the LQCD data in each of the two channels. The results are shown in Table~\ref{table:pheno_fit}, where we also list the resulting values of $\delta f_1(0)$ at the physical point and where we compare with the AGt-based fits done in Ref.~\cite{Sasaki:2012ne}. As we can see, the corrections to $f_1(0)$ at $\mathcal{O}(p^5)$ needed to fit the LQCD data of Ref.~\cite{Sasaki:2012ne} would be $\sim-7\%$ and $\sim-6\%$ for the $\Sigma N$ and the $\Xi\Sigma$ channels, respectively. These corrections are larger than those one would expect from $\mathcal{O}(m_s^2)$ terms and in this scenario one will certainly conclude that the chiral expansion for $f_1(0)$ is very slow- or non-converging. Further LQCD simulations at lighter quark masses and with full control of systematic uncertainties will be very helpful to clarify this issue.

\begin{figure}[t]
\centerline{\includegraphics[scale=0.45]{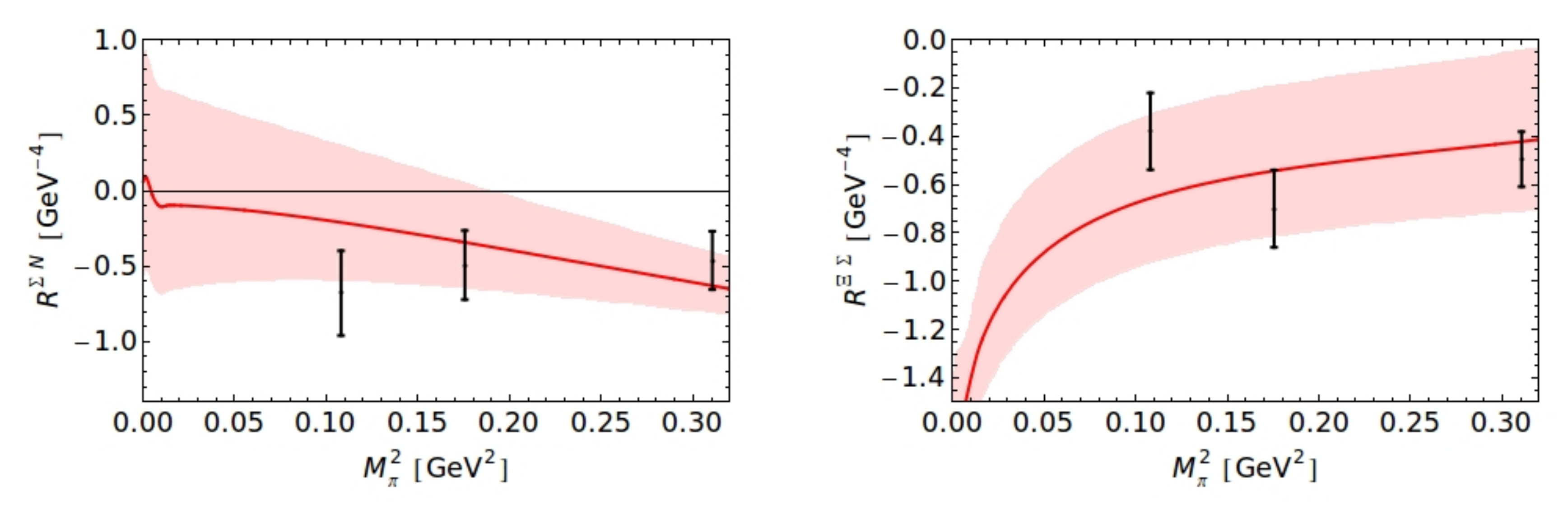}}
\caption{SU(3)-breaking corrections to $f_1(0)$ in infinite volume as functions of the pion mass, $m_\pi$, in covariant BChPT. An analytical $\mathcal{O}(p^5)$ term has been added to the full $\mathcal{O}(p^4)$ BChPT results.\label{fig:plotRs}}
\end{figure}
This exercise is also illustrative in highlighting the role that chiral dynamics can play in the SU(3)-breaking of $f_1(0)$. As shown in Table~\ref{table:pheno_fit}, the results of the
BChPT-inspired fits are quite different to those based on the AGt, where one fits a term like in Eq.~(\ref{eq:AGtf1}) ignoring the effects of the chiral loops. The differences are just a consequence of the structure of the chiral expansion of $f_1(0)$ discussed in Sec.~\ref{sec:BChPT}. Indeed, the loop corrections do not scale quadratically in $m_s$, but like $\mathcal{O}(m_s^{1/2})$, $\mathcal{O}(m_s)$, etc., as soon as $m_s$ becomes much larger than $m_{ud}$ approaching the physical point. To study better the impact these effects can have on the chiral extrapolation of $f_1(0)$ we define the following function~\cite{Guadagnoli:2006gj,Sasaki:2008ha,Sasaki:2012ne}:
\begin{equation}
R=\frac{\delta f_1(0)}{(m_K^2-m_\pi^2)^2},
\end{equation}
where we have factored out a dependence $\sim m_s^2$ from $\delta f_1(0)$. In Fig.~\ref{fig:plotRs} we show the results of our BChPT-inspired fits compared against the LQCD results. As one can see, the chiral behavior predicted by BChPT is very different from the constant dependence expected by the AGt and terms which are clearly nonanalytical in $m_q$ dominate the extrapolation around the physical point. In fact the results in the extrapolation can be very different if one account for these effects using the BChPT calculation discussed in this work.

\begin{figure}[t]
\centerline{\includegraphics[scale=1.0,angle=0]{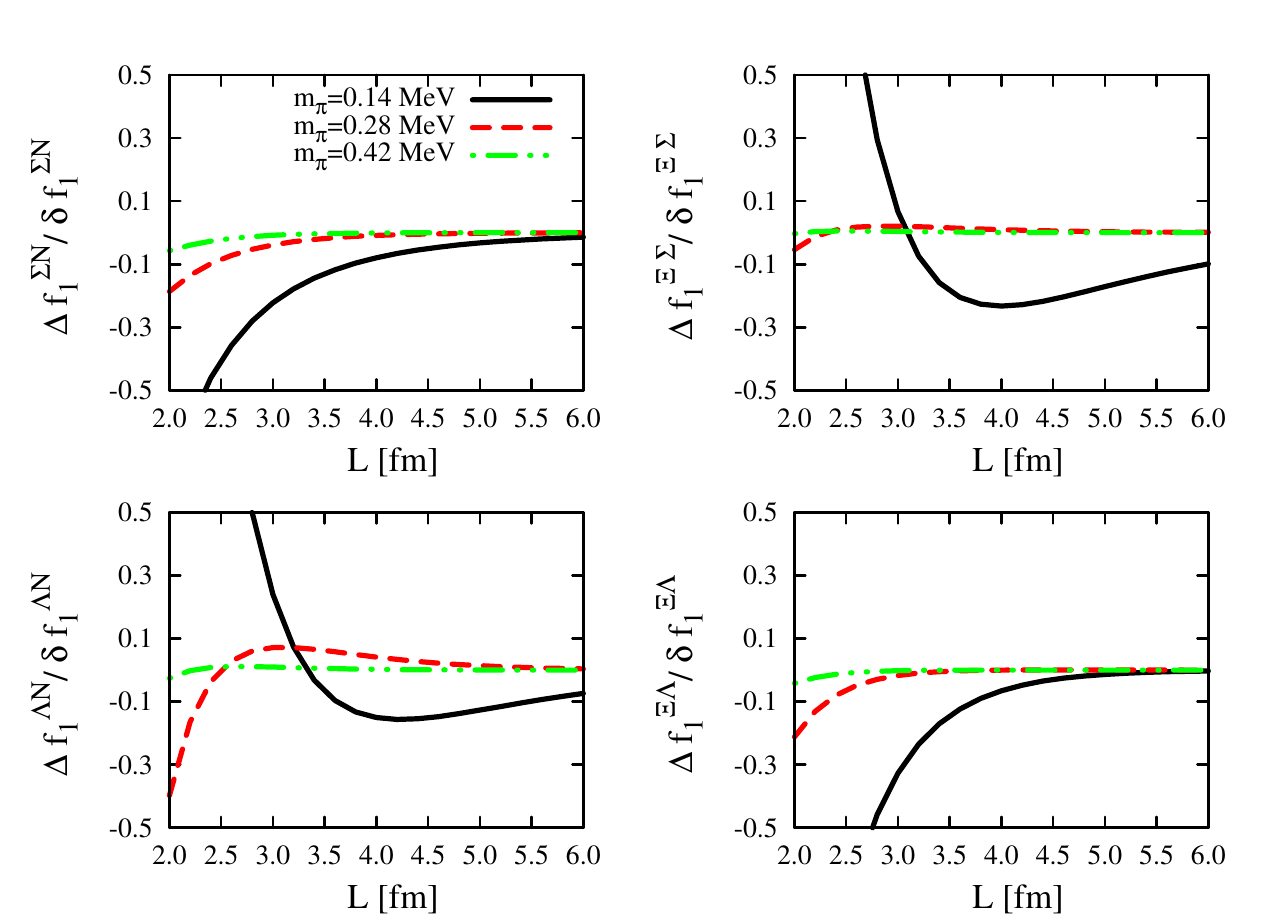}}
\caption{Ratio of finite-volume corrections to SU(3)-breaking corrections to $f_1(0)$ as a function of $L$ and $m_\pi$ (see text for details).\label{fig:FV}}
\end{figure}

\subsection{A close-up onto the volume dependence}

Although FVCs are small in the current LQCD setup of Ref.~\cite{Sasaki:2012ne}, they can become large with decreasing $m_\pi$. In Fig.~\ref{fig:FV}, we show the ratio of the FVCs, $\Delta f_1(0)$, to the corresponding SU(3)-breaking corrections, $\delta f_1(0)$, as a function of the box size $L$ for different $m_\pi$ and with the strange quark mass set at its physical value. The $\mathcal{O}(p^2)$ octet baryon masses appearing in the calculation are determined using the LECs obtained in Ref.~\cite{Ren:2012aj} by fitting to the available LQCD data and the $\mathcal{O}(p^2)$ decuplet baryon  masses are determined using the LECs given in the text. Because at $m_\pi=0.14$ GeV, $\delta f_1^{\Xi\Sigma}=0.2\%$ is accidentally small, we have multiplied the corresponding $\delta f_1$ by a factor of 3 to calculate the ratio.

One can clearly see that at $m_\pi=0.14$ GeV, with $L\approx3 $ fm as in Ref.~\cite{Sasaki:2012ne}, FVCs can be as large as 30\%.  In addition, in the $\Xi\Sigma$ and $\Lambda N$ channels,  non-monotonous change of FVCs with $L$ is observed. It seems that for LQCD simulations with light-quark masses close to their physical values, a box size of 5 to 6 fm would be necessary to keep FVCs smaller than 10\%. The calculations presented in this work could be used in the future for testing and correcting the finite volume effects in future LQCD calculations.

\section{Summary}

We have studied the discrepancy between  the latest $n_f=2+1$ LQCD simulation results on the hyperon vector couplings, $f_1(0)$, and the corresponding predictions of covariant baryon chiral perturbation up to $\mathcal{O}(p^4)$.  In particular, we studied  the chiral extrapolation of and finite-volume corrections to the LQCD data. Our studies showed that at the present LQCD setup, finite-volume corrections are small and can be safely neglected. Furthermore, non-analytical chiral contributions can become important in extrapolating LQCD results to the physical point, while a naive application of the Ademollo Gatto theorem could be inappropiate.  Nevertheless, our studies showed that none of the above two effects can explain the discrepancy between current fully dynamical LQCD simulations and the BChPT predictions without sizable $\mathcal{O}(p^5)$ contributions. Clearly, further studies, particularly, LQCD simulations with smaller light-quark masses and larger volumes, are needed to clarify the situation.

\section{Acknowledgements}
This work was partly supported by the National
Natural Science Foundation of China under Grant No. 11375024 and  the
New Century Excellent Talents in University Program of Ministry of Education of China under
Grant No. NCET-10-0029. J.M.C has received funding from the People Programme (Marie Curie Actions) of the European Union's Seventh Framework Programme (FP7/2007-2013) under REA grant agreement n PIOF-GA-2012-330458 and acknowledges the Spanish Ministerio de Econom\'ia y Competitividad and European FEDER funds under the contract FIS2011-28853-C02-01 for support.

\section{Appendix: Finite-volume corrections to $f_1(0)$ at $\mathcal{O}(p^3)$}

 We introduce the following notations for FVCs
\begin{equation}
\Delta G_L= G_L-G_\infty,
\end{equation}
where $G$ denote a generic loop integral and $L(\infty)$ denotes the corresponding result in finite volume (infinite space-time).

The $\mathcal{O}(p^3)$ results introduced by virtual octet baryons have the following
structure for the transition $i\rightarrow j$:
\begin{eqnarray}\label{eq:sumo3}
 \Delta \delta_B^{(2)}(i\rightarrow j)&=&\sum_{M=\pi,\eta,K}\beta^\mathrm{BP}_M \Delta H_\mathrm{BP}(m_M)+\sum_{M=\pi,\eta} \beta^\mathrm{MP}_{M} \Delta H_\mathrm{MP}(m_M,m_K)+\sum_{M=\pi,\eta,K} \beta^\mathrm{KR}_M \Delta H_\mathrm{KR}(m_M)\nonumber\\
&&-\frac{3}{8}\sum_{M=\pi,\eta}\Delta H_\mathrm{TD1}(m_M,m_K)+\frac{3}{8}\sum_{M=\pi,\eta}\Delta H_\mathrm{TD2}(m_M)
+\frac{3}{4}\Delta H_\mathrm{TD2}(m_K)\nonumber\\
&&+\frac{1}{2}\sum_{M=\pi,\eta,K}(\beta^\mathrm{WF}_M(i)+\beta^\mathrm{WF}_M(j))\Delta H_\mathrm{WF}(m_M),
\end{eqnarray}
where $\beta^\mathrm{BP}$, $\beta^\mathrm{MP}$, $\beta^\mathrm{KR}$, and $\beta^\mathrm{WF}$ are given in the Appendix of Ref.~\cite{Geng:2009ik}, and the FVCs $\Delta H_\mathrm{BP}$, $\Delta H_\mathrm{MP}$, $\Delta H_\mathrm{KR}$, $\Delta  H_\mathrm{TD1}$, $\Delta H_\mathrm{TD2}$, and $
\Delta H_\mathrm{WF}$  are, respectively,
\begin{eqnarray}
\Delta H_\mathrm{BP}&=&\frac{-1}{F_0^2}\int^1_0 dx\, (1-x)\left\{
\delta_{1/2}(\mathcal{M}_\mathrm{BP}^2)+\frac{1}{8} \left(-12 (-2+x) x m_0^2-8 \mathcal{M}_\mathrm{BP}^2 \right) \delta_{3/2}(\mathcal{M}_\mathrm{BP}^2) \right.\\
&& \left.\hspace{2cm} +\frac{1}{8} \left(3 (-1+x)^4 m_0^4-6 \left(1+2 x-x^2\right) m_0^2 \mathcal{M}_\mathrm{BP}^2+3 \mathcal{M}_\mathrm{BP}^4\right) \delta_{5/2}(\mathcal{M}_\mathrm{BP}^2) \right\}, \nonumber
\end{eqnarray}

\begin{eqnarray}
\Delta H_\mathrm{MP}&=&\frac{-1}{F_0^2}\int^1_0 dx \int^{1-x}_0 dy\, \left\{ \delta_{1/2}(\mathcal{M}_\mathrm{MP}^2)+\frac{1}{4} \left(-3 x (2+3 x) m_0^2-\mathcal{M}_\mathrm{MP}^2\right)\delta_{3/2}(\mathcal{M}_\mathrm{BP}^2) \right.\nonumber\\
&& \left.\hspace{3cm} +\frac{1}{4} \left(3 x^4 m_0^4+3 x (2+x) m_0^2 \mathcal{M}_\mathrm{MP}^2 \right) \delta_{5/2}(\mathcal{M}_\mathrm{MP}^2) \right\},
\end{eqnarray}

\begin{equation}
\Delta H_\mathrm{KR}=\frac{-1}{F_0^2}\int^1_0 dx\,\frac{1}{2} \left\{ \delta_{1/2}(\mathcal{M}_\mathrm{KR}^2) -\frac{1}{2} \left((x-1)^2 m_0^2+\mathcal{M}_\mathrm{KR}^2\right) \delta_{3/2}(\mathcal{M}_\mathrm{KR}^2) \right\},
\end{equation}

\begin{equation}
\Delta H_\mathrm{TD1}=\frac{-1}{F_0^2}\int^1_0 dx\, \delta_{1/2}(\mathcal{M}_\mathrm{TD1}^2),
\end{equation}

\begin{equation}
\Delta H_\mathrm{TD1}=\frac{-1}{F_0^2}\int^1_0 dx\, \frac{1}{2} \delta_{1/2}(\mathcal{M}_\mathrm{TD2}^2),
\end{equation}

\begin{eqnarray}
\Delta H_\mathrm{WF}&=&\frac{-1}{F_0^2}\int^1_0 dx\, \left\{ x \delta_{1/2}(\mathcal{M}_\mathrm{WF}^2)+\frac{1}{4} \left(x \left(6+x-9 x^2\right) m_0^2+(-1-x) \mathcal{M}_\mathrm{WF}^2\right) \delta_{3/2}(\mathcal{M}_\mathrm{WF}^2) \right.\\
&& \left. +\frac{1}{4} \left(3 (-1+x)^2 x^3 m_0^4+3 (-1+x) x m_0^2 \left(m_M^2 (-1+x)+(3+x) \mathcal{M}_\mathrm{WF}^2\right)\right) \delta_{5/2}(\mathcal{M}_\mathrm{WF}^2) \right\},
\nonumber
\end{eqnarray}
where $m_0$  is the chiral limit octet baryon mass,  and $m_1$, $m_2$, and $m_M$ are the relevant meson masses.

The $\mathcal{O}(p^3)$ results induced by virtual decuplet baryons are
\begin{eqnarray}\label{eq:sumd3}
 \Delta \delta_D^{(2)}(i\rightarrow j)&=&\sum_{M=\pi,\eta,K}\gamma^\mathrm{BP}_M \Delta D_\mathrm{BP}(m_M)+\sum_{M=\pi,\eta} \gamma^\mathrm{MP}_{M} \Delta D_\mathrm{MP}(m_M,m_K)+\sum_{M=\pi,\eta,K} \gamma^\mathrm{KR}_M \Delta D_\mathrm{KR}(m_M)\nonumber\\
&&+\frac{1}{2}\sum_{M=\pi,\eta,K}(\gamma^\mathrm{WF}_M(i)+\gamma^\mathrm{WF}_M(j)) \Delta D_\mathrm{WF}(m_M),
\end{eqnarray}
where $\gamma^\mathrm{BP}$, $\gamma^\mathrm{MP}$, $\gamma^\mathrm{KR}$, and $\gamma^\mathrm{WF}$ are given in the Appendix of Ref.~\cite{Geng:2009ik}, and the FVCs
$\Delta D_\mathrm{BP}$, $\Delta D_\mathrm{MP}$, $\Delta D_\mathrm{KR}$,  and $
\Delta D_\mathrm{WF}$  are, respectively,

\begin{eqnarray}
\Delta D_\mathrm{BP}&=&\frac{C^2}{F_0^2 m_D^2} \int^1_0 dx\, \left\{   \frac{1}{2}   \left(3 (1+x)^2 m_0^2-5 \mathcal{\tilde{M}}_\mathrm{BP}^2+6 (1+x) m_0 \Delta _D+3 \Delta _D^2\right)\delta_{3/2}(\mathcal{\tilde{M}}_\mathrm{BP}^2)\right.\nonumber\\
&&\left. +\frac{1}{2}   \left(-3 (1+x)^2 m_0^2 +3 \mathcal{\tilde{M}}_\mathrm{BP}^2-6 (1+x) m_0  \Delta _D-3  \Delta _D^2\right) \mathcal{\tilde{M}}_\mathrm{BP}^2\delta_{5/2}(\mathcal{\tilde{M}}_\mathrm{BP}^2)\right.\nonumber\\
&&+\left. \delta_{1/2}(\mathcal{\tilde{M}}_\mathrm{BP}^2) \right\}\frac{-m_0^2}{6} (1-x),
\end{eqnarray}

\begin{eqnarray}
\Delta D_\mathrm{MP}&=&\frac{C^2}{F_0^2 m_D^2}\int^1_0 dx \int^{1-x}_0 dy \,  \left\{ \left(3 (-2+x) x m_0^2+\mathcal{\tilde{M}}_\mathrm{MP}^2-3 x m_0 \Delta _D\right)\delta_{3/2}(\mathcal{\tilde{M}}_\mathrm{MP}^2)+\right.\nonumber\\
&& \left. \quad \quad    \left(-3 (-2+x) x m_0^2 +3 x m_0  \Delta _D\right) \mathcal{\tilde{M}}_\mathrm{MP}^2 \delta_{5/2}(\mathcal{\tilde{M}}_\mathrm{MP}^2) -  \delta_{1/2}(\mathcal{\tilde{M}}_\mathrm{MP}^2)\right\}  \frac{m_0^2}{6},
\end{eqnarray}

\begin{equation}
\Delta D_\mathrm{KR}=\frac{C^2}{F_0^2 m_D^2}\int^1_0 dx\,\frac{m_0}{6} \left\{ \left((1+x) m_0+\Delta _D\right)
\left(\delta_{1/2}(\mathcal{\tilde{M}}_\mathrm{KR}^2)-\mathcal{\tilde{M}}_\mathrm{KR}^2
\delta_{3/2}(\mathcal{\tilde{M}}_\mathrm{KR}^2)\right)
\right\},
\end{equation}

\begin{eqnarray}
\Delta D_\mathrm{WF} &=&  \frac{C^2}{F_0^2 m_D^2}\int^1_0 dx\, \frac{m_0}{6}  \left\{ \left(3 m_0^2 (x-1) x m_D-3 m_0^3  (x-1)^2 x\right)\mathcal{\tilde{M}}_\mathrm{WF}^2 \delta_{5/2}(\mathcal{\tilde{M}}_\mathrm{WF}^2)+ \right.\nonumber\\
&&\left.
\left(-2 \mathcal{\tilde{M}}_\mathrm{WF}^2 m_D-3 m_0^2 (x-1) x m_D+3 m_0 \mathcal{\tilde{M}}_\mathrm{WF}^2 (x-1)+3 m_0^3 (x-1)^2 x\right)\delta_{3/2}(\mathcal{\tilde{M}}_\mathrm{WF}^2)\right.\nonumber\\
&&\left.+  \left(2 m_D-3 m_0 (x-1)\right)\delta_{1/2}(\mathcal{\tilde{M}}_\mathrm{WF}^2)\right\},
\end{eqnarray}
where $m_D$ is the chiral limit decuptet baryon mass, and $\Delta_D=m_D-m_0$.

In the above equations, the master formulas $\delta_r(\mathcal{M}^2)$ are defined as
\begin{equation}
\delta_r(\mathcal{M}^2)=\frac{2^{-1/2-r}(\sqrt{\mathcal{M}^2})^{3-2r}}{\pi^{3/2}\Gamma ( r )}\sum\limits_{\vec{n}\ne0} (L \sqrt{\mathcal{M}^2} |\vec{n}|)^{-3/2+r} K_{3/2-r}(L \sqrt{\mathcal{M}^2}|\vec{n}|),
\end{equation}
where $K_n(z)$ is the modified Bessel function of the second kind, and $\sum\limits_{\vec{n}\ne0}\equiv \sum\limits_{n_x=-\infty}^{+\infty}\sum\limits_{n_y=-\infty}^{+\infty}\sum\limits_{n_z=-\infty}^{+\infty}(1-\delta(|\vec{n}|,0))$. The $\mathcal{M}^2$ for different diagrams are defined as
\begin{equation}
\mathcal{M}_\mathrm{BP}^2=m_0^2(1-x)^2+m_M^2 x - i \epsilon,
\end{equation}
\begin{equation}
\mathcal{M}_\mathrm{MP}^2=m_0^2 x^2 + m_1^2 y - m_2^2 (x+y-1)- i \epsilon,
\end{equation}
\begin{equation}
\mathcal{M}_\mathrm{KR}^2=m_0^2(1-x)^2+ m_M^2 x -i \epsilon,
\end{equation}
\begin{equation}
\mathcal{M}_\mathrm{TD1}^2= m_1^2 x +m_2^2 (1-x)-i\epsilon,
\end{equation}
\begin{equation}
\mathcal{M}_\mathrm{TD2}^2=m_M^2-i\epsilon,
\end{equation}
\begin{equation}
\mathcal{M}_\mathrm{WF}^2=m_0^2 x^2 + m_M^2(1-x)-i\epsilon,
\end{equation}
\begin{equation}
\mathcal{\tilde{M}}_\mathrm{BP}=m_D^2(1-x)+x m_M^2+m_0^2 x (x-1) -i\epsilon,
\end{equation}
\begin{equation}
\mathcal{\tilde{M}}_\mathrm{MP}=m_D^2 x + m_0^2x(x-1) +m_1^2 y - m_2^2 (x+y-1) -i\epsilon,
\end{equation}
\begin{equation}
\mathcal{\tilde{M}}_\mathrm{KR}=m_D^2 (1-x) + xm_M^2+ m_0^2x(x-1)  -i\epsilon,
\end{equation}
\begin{equation}
\mathcal{\tilde{M}}_\mathrm{WF}=m_D^2 x + m_M^2 (1-x)+ m_0^2 x (x-1)  -i\epsilon.
\end{equation}

\end{document}